\documentclass{article}  
\usepackage[utf8]{inputenc}  
\usepackage[total={18cm, 24 cm}, centering]{geometry}  
\usepackage[authoryear,round, square,numbers]{natbib}
\setcitestyle{square} 
\usepackage{authblk} 
\usepackage{url}
\usepackage{graphicx} 
\usepackage[export]{adjustbox} 
\usepackage{letltxmacro}
\usepackage[dvipsnames]{xcolor} 
\usepackage{booktabs} 
\usepackage{float} 
\usepackage{titlesec} 

\title{\textbf{Machine Learning Analysis of the Impact of Increasing the Minimum Wage on Income Inequality in Spain from 2001 to 2021}}
\author[1]{Marcos Lacasa Cazcarra}
\affil[1]{Education Department, Catalonia Goverment}
\date{}  

\begin{document}
\maketitle
\begin{abstract}
    This paper analyzes the impact of the National Minimum Wage from 2001 to 2021. The MNW increased from €505.7/month (2001) to €1,108.3/month (2021). Using the data provided by the Spanish Tax Administration Agency, databases that represent the entire population studied can be analyzed. More accurate results and more efficient predictive models are provided by these counts. This work is characterized by the database used, which is a national census and not a sample or projection. Therefore, the study reflects results and analyses based on historical data from the Spanish Salary Census 2001-2021. Various machine-learning models show that income inequality has been reduced by raising the minimum wage. Raising the minimum wage has not led to inflation or increased unemployment. On the contrary, it has been consistent with increased net employment, contained prices, and increased corporate profit margins.

    The most important conclusion is that an increase in the minimum wage in the period analyzed has led to an increase in the wealth of the country, increasing employment and company profits, and is postulated, under the conditions analyzed, as an effective method for the redistribution of wealth.
\end{abstract}
\section{Introduction}

    The economic literature has analyzed the impact of the minimum wage on employment, prices and growth. The idea of the neo-liberal trend is that higher wages mean higher costs for the producers. If this is associated with increased productivity, it should not be associated with increased prices. To do otherwise would lead to an inflationary spiral. The first studies in the United States concluded that workers affected by an increase in the National Minimum Wage (NMW) are those who earn a wage below the NMW \cite{Lopresti2016-ov}. For this group of workers, the increase in the NMW increases wages but reduces hours worked \cite{Neumark2004-uv}. In contrast, similar analyses in other countries, such as Germany, come to different conclusions \cite{Dustmann2021-fa} and an increase in the NMW does not alter the demand for employment and causes a shift in the demand for jobs that is perceived as better by workers with wages similar to the NMW. It is not easy to affirm any of these positions, and cautious postulates must be considered \cite{Clemens2021-vn}. Previous studies have linked increases in the NMW to inflation \cite{Sellekaerts1982-wh}. In a historical series in Turkey, the increase in the NMW has been associated with an increase in prices and unemployment \cite{Kemal2019-yg}.

    The data used in the works analyzed are based on samples, surveys or statistical projections based on economic models. It is considered important to comment on a report by the Spanish Central Bank which analyzes the consequences of an increase in the NMW from €858/month in 2018 to €1050/month in 2019 \cite{Lacuesta_Gabarain2019-cg}. This paper has considered for its analysis the data of the Employment History Report Sample between 2013-2017 (MCVL - Portal Estadisticas - Seguridad Social, n.d.). Based on the projected data, young people (under 25) and people over 45 have more than a 20\% risk of losing their jobs as a result of the minimum wage increase. These findings are in contrast to publications that conclude that there is insufficient evidence of a relationship between minimum wage increases and unemployment \cite{Neumark2022-hj}.

    There are significant differences between the minimum wages of the EU member states \\ (source: Eurostat; EARN\_Dist\_MW\_Dist\_CUR). The first group would be formed by Germany, the Netherlands, Belgium, Ireland and France and the range would be €1750-€2000/month. Spain and Slovenia would form a second group with a similar minimum wage of around €1250/month and the third group would be formed by the rest of the EU member states with a MNW up to €1000/month.

    MNW definition in Spain is a political decision. Particularly in recent years, the most progressive governments have led to significant increases. This paper analyzes the impact of the MNW from 2001 to 2021. The MNW increased from €505.7/month (2001) to €1,108.3/month (2021). This work is characterized by the database used, which is a national census and not a sample or projection. Therefore, the study reflects results and analyses based on historical data from the Spanish Salary Census 2001-2021.

\section{Methods}
\subsection{The database}

The Spanish Tax Administration Agency (Spanish: Agencia Estatal de Administración Tributaria, AEAT) is a public institution. It is attached to the Ministry of Economy and Finance through the former State Secretariat for Finance and Budget. It handles the application of the tax system under the constitutional principle that everyone must contribute to supporting public expenditure according to their economic capacity. It prepares the annual tax collection reports, which provide information on the amount and annual evolution of the tax revenues managed by the AEAT. It offers a database in Excel format called "Distribucion salarios" in the supplementary material. This database collects information on employees in Spain who receive income from a company or entity required to provide a list of those receiving such income. This database does not include information on households that pay wages to employees in the household, nor does it reflect the reported wages of the self-employed.

Payments are defined as income, in cash or kind, paid by the reporting unit (company or institution) in the form of annual income. Employees are included in the database even if they worked for only one day. If an employee has worked for more than one company or entity, the amount shown is the sum of all payments made to that employee by the different companies from which he/she received his/her salary. The information is completely anonymous and is presented in intervals of \$200, from \$0 to \$80,000. The last interval is open, with no maximum value specified. A total of 400 annual salary levels are presented. Each cell represents the value corresponding to each bracket by year (2001-2021). Three variables are analyzed: number of employees, salary, and withholding tax.

\subsection{Additional Macroeconomic Data}

This paper analyzes other macroeconomic data regularly published by the National Statistics Institute (INE), such as; MNW, Consumer Price Index, Unemployment Index, Gross Domestic Product, and Public Debt. In no case are the data deflated, as they are all nominal.

\subsection{Calculation of the Gini Index}
The formula used to calculate the Gini index is:
$$G= \frac{x}{n^2 \bar{x}}\sum^n_{i=1} i(x_i - \bar{x})$$

To calculate the Gini index for a year, the number of registered employees in that year is needed. The database provides this information in brackets. The number of employees with a gross annual income equal to the interval, the total gross income, and the total income tax withheld are provided for each interval (range of \$200). The mean is considered to be the distribution value that makes the variance zero since the probability distribution of each ith interval is considered to be identically distributed. This results in a vector for each interval composed of j values that are identical and equal to the mean of that interval.

$$Vector\: i^{th} : n^o\: employees(j)\: \bar{x_i}$$

Each year vector is a union of all intervals vectors of this year.

\subsection{Graph analysis}

To analyze the relationships between macroeconomic variables, graph theory was used. A graph is a collection of nodes (also called vertices) connected by edges (undirected)\cite{Newman2018-gc}. The pattern of interactions between the nodes (individuals or entities) can be captured through the graph structure. The purpose of graph (or network) analysis is the study of relationships between individuals to discover knowledge about global and local structures.

In this paper, the nodes of the graph are defined as all macroeconomic variables, and the edges are defined as moderate or strong correlations between them. The linear correlation between two nodes is represented by $corr(i,j)$, and the Spearman correlation is defined as moderate or strong if $corr(i,j) \ge 0.5$ \cite{Suchowski_undated-oo} in case of direct correlation. An $edge(i,j)$  is defined if  $abs(corr(i,j) \ge 0.5$.

Detecting communities in networks is one of the most popular topics in modern network science. Communities, or clusters, are typically groups of nodes that are more likely to be connected than to members of other groups, although other patterns are possible. There are no universal protocols, neither for defining a community itself nor for other crucial issues such as validating algorithms and comparing their performance.

The Louvain method hierarchically performs a greedy optimization \cite{Blondel2008-ax}, assigning each vertex to the community of its neighbors that yields the highest number, and creating a smaller weighted network whose vertices are the clusters found previously \cite{Fortunato2016-yw}. Partitions found on this super-network hence consist of clusters including the ones found earlier, and represent a higher hierarchical level of clustering. Software used: Gephi v 0.10 \cite{Jacomy2014-oe}.

\subsection{Multivariate Linear Regression}
The statistical model is assumed to be $$Y = X \beta + \mu$$
 where  $\mu\ N (0, \mathrm{\Sigma})$.
The ordinary least squares for independent identically distributed errors (MSE).
R- square formula is,

$$R^2=1-\frac{\mathrm{sum\ squared\ regression\ (SSR)}\ }{\mathrm{total\ sum\ of\ squares\ (SST)}}\ =1-\frac{\sum\left(y_i\ -\hat{y}\ \right)^2\ }{\sum\left(y_i-\ \bar{y}\right)^2\ }$$

The Durbin-Watson (DW) statistic is a test for autocorrelation in the residuals from a statistical model or regression analysis. Values from 0 to less than 2 indicate positive autocorrelation and values from 2 to 4 indicate negative autocorrelation. 

Durbin-Watson test statistics d is given as,
$$d\ =\ \frac{\sum_{i\ =\ 2}^{N}\left(e_i-e_{i-1}\right)^2}{\sum_{i\ =\ 1}^{N}{e_i}^2}$$

where $N$ is the number of observations and $e_i$ is the residual for each observation $(i)$.
The software used is statsmodels packages for Python\cite{Seabold2010-xa}.

\subsection{Random Forest Regressor}

The Random Forest Regressor (RFR) is an ensemble learning model. It combines the predictions of multiple models to produce more accurate results than a single model \cite{Breiman2001-ea}. A decision tree (DT) is a simple model that predicts the outcome by performing a partition based on the predictor (input variable) that provides the greatest reduction in mean squared error (MSE).
$$MSE\ =\ \frac{1}{n}\sum^n_i = (y_i-\hat{y})^2$$

Where $y$ and $\hat{y}$  are the measured and predicted values of the samples in a node, respectively, and n is the number of samples in a node. A node that cannot branch further due to a non-decreasing MSE is called a leaf node, and the average of the samples in that node becomes a candidate for prediction. When unseen data is entered into the final DT model, the data moves according to predetermined branching criteria. The value of the leaf node where the data finally arrives is used as the predicted value of the DT. Scitik-Learn is de machine learning software used in Python\cite{Pedregosa2012-ts}.

\section{Results}
\subsection{Calculation of the Gini Index}

The Gini index of the gross wages received by all workers in the years under study is analyzed and called Gross-Gini. A similar analysis called the net Gini, is carried out on net income (gross salary minus withholding tax). It shows the effect of income tax progressivity as measured by the difference between the two indexes.\textbf{Table \ref{table:table1}} shows the annual results of the database “Distribucion salarios” provided by AEAT. The nominal increase in average annual gross earnings was like the annual increase in minimum wage. Over the twenty years of the study, they increased by €7,500 and €7,300, respectively. The annual minimum wage percentile is calculated concerning annual gross income. In most years, more than 30\% of workers did not earn the equivalent of the annual minimum wage. This is due to a significant level of underemployment. Tourism in Spain is seasonal and accounts for more than 13\% of the total employed \cite{Cabrer-Borras2021-sh}. Seasonality in the agricultural sector in Spain accounts for 5\% of employment \cite{Molinero-Gerbeau2021-si}. Over the period under review, the progressivity of income tax has remained stable. Income inequality is favored by this progressivity. The difference between the calculated Gini indices indicates the reduction in income inequality brought about by the progressivity of the tax.

\begin{figure}
    \centering
    \caption{Change in average withholding tax by income ranges}
    \includegraphics[width = 18cm]{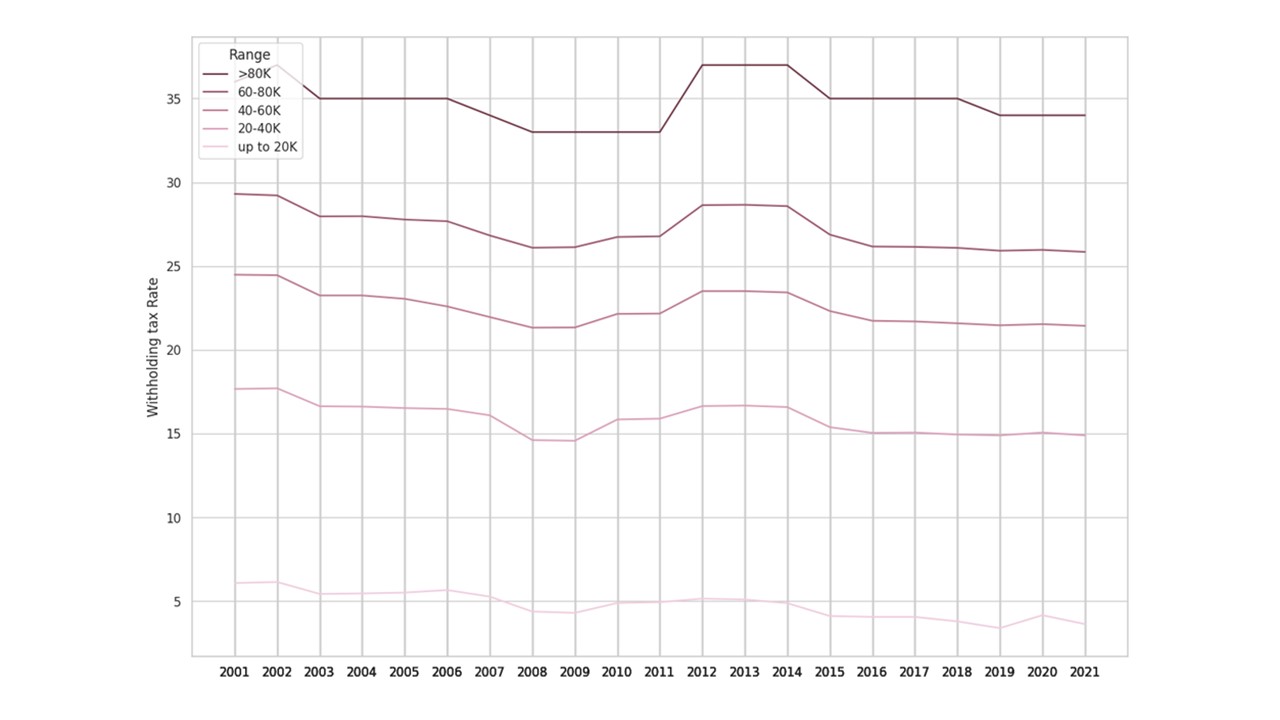}
    \label{fig:figure1}
    \medskip
    \small\raggedright \textit{
    Five intervals of €20,000 from the database are defined. The lines show the change in the average withhold rate over time for each.}
\end{figure}

\begin{table}[ht] 
    \caption{Macroeconomics time series}
    \centering
    \begin{tabular}{lllllll}\\
    \toprule

years & Mean Gross Salary & NMW        & Percentile & Gross-Gini & Average rate of    taxation & Net-Gini \\
\hline
2001  & €13,932.00        & €6,068.40  & 28.4       & 0.465      & 14.3\%                      & 0.424    \\
2002  & €14,370.00        & €6,190.80  & 28.3       & 0.462      & 14.7\%                      & 0.421    \\
2003  & €14,963.00        & €6,316.80  & 28.2       & 0.463      & 14.0\%                      & 0.423    \\
2004  & €15,658.00        & €6,447.60  & 26.5       & 0.457      & 14.4\%                      & 0.416    \\
2005  & €16,018.00        & €7,182.00  & 29.3       & 0.462      & 14.7\%                      & 0.422    \\
2006  & €16,849.00        & €7,573.20  & 28.8       & 0.459      & 15.1\%                      & 0.418    \\
2007  & €18,087.00        & €7,988.40  & 27.0       & 0.448      & 15.1\%                      & 0.406    \\
2008  & €18,996.00        & €8,400.00  & 27.8       & 0.453      & 14.5\%                      & 0.411    \\
2009  & €19,085.00        & €8,736.00  & 30.8       & 0.466      & 14.8\%                      & 0.424    \\
2010  & €19,113.00        & €8,866.80  & 31.0       & 0.467      & 15.8\%                      & 0.426    \\
2011  & €19,102.00        & €8,979.60  & 32.0       & 0.469      & 15.9\%                      & 0.428    \\
2012  & €18,601.00        & €8,979.60  & 33.0       & 0.472      & 16.6\%                      & 0.427    \\
2013  & €18,505.00        & €9,034.80  & 34.3       & 0.482      & 16.7\%                      & 0.437    \\
2014  & €18,420.00        & €9,034.80  & 34.7       & 0.483      & 16.6\%                      & 0.439    \\
2015  & €18,645.00        & €9,080.40  & 34.1       & 0.483      & 15.6\%                      & 0.440    \\
2016  & €18,835.00        & €9,172.80  & 34.1       & 0.483      & 15.3\%                      & 0.439    \\
2017  & €19,172.00        & €9,908.40  & 34.6       & 0.472      & 15.3\%                      & 0.428    \\
2018  & €19,809.00        & €10,303.20 & 33.8       & 0.465      & 15.4\%                      & 0.419    \\
2019  & €20,566.00        & €12,600.00 & 37.4       & 0.457      & 15.3\%                      & 0.410    \\
2020  & €20,503.00        & €13,299.60 & 43.1       & 0.474      & 16.0\%                      & 0.429    \\
2021  & €21,519.00        & €13,299.60 & 38.1       & 0.458      & 16.0\%                      & 0.412    \\
    \bottomrule
    \\
    \end{tabular}

    \small\raggedright
    \textit{
\textbf{Mean Gross Salary}. - annual average income per employee. \textbf{NMW}. -  National minimum wage per year. \textbf{Percentile}. -  Annual minimum wage percentile of total gross annual income. \textbf{Gross-Gini}. - The Gini index is calculated from the vector of all gross incomes per year. \textbf{Average rate of taxation}. -  Ratio of the total amount withheld to the total annual income. \textbf{Net-Gini}. - The Gini index is calculated from the vector of all net incomes (Gross income minus withholdings) per year.}
\label{table:table1}
\end{table}
In the following analysis of the dataset, we reduce the intervals from 400 to 5. To do this, we will increase the intervals from €200 to €20,000. The average withholding tax per interval and its evolution are analyzed over the study period in \textbf{Figure \ref{fig:figure1}}. The average withholding rate shows a general downward trend that is more pronounced in the lower average wage brackets. Comparing the ratio from 2001 to 2021, the decline is similar in the five three-percentage-point intervals, but the impact on net income is much greater in the range below 20K (from 6.09\% in 2001 to 3.64\% in 2021) than in the range between 60K and 80K (29.31\% in 2001 to 25.85\% in 2021).

\subsection{Mean Salary analysis}

\begin{figure}[H]
    \centering
    \caption{Evolution of Average Gross Wage vs. GDP}
    \includegraphics[width = 18cm]{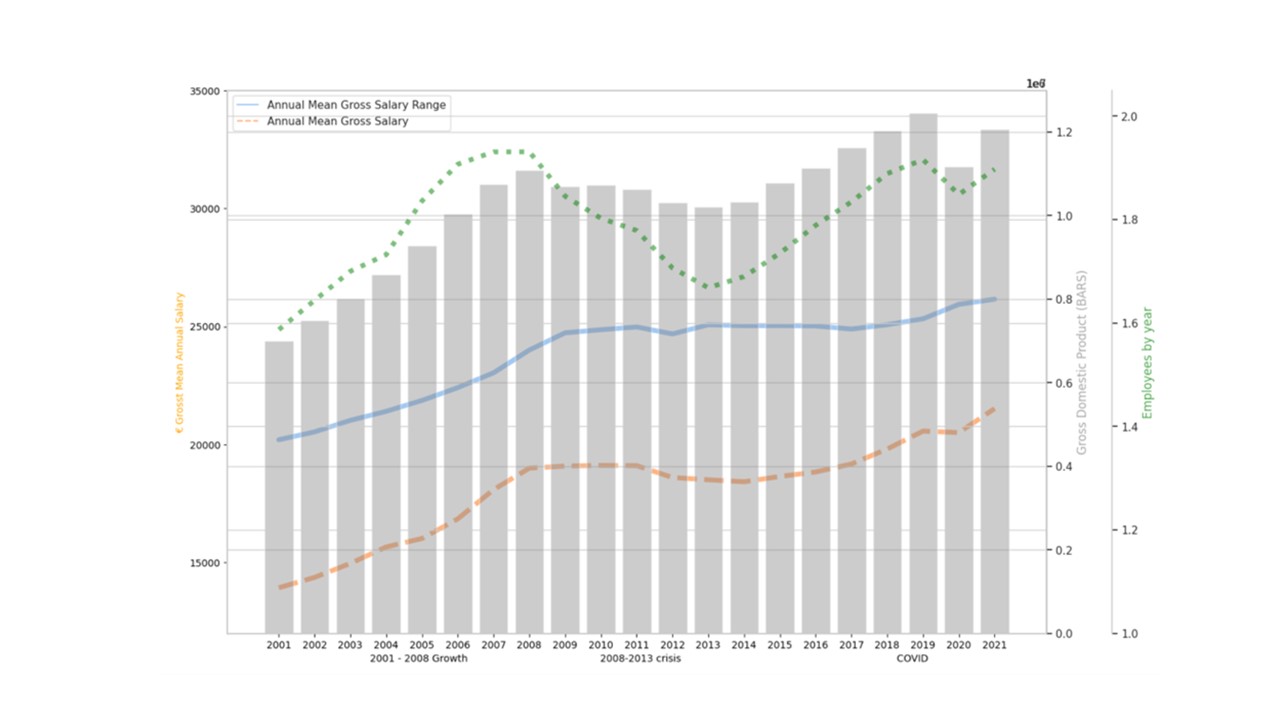}
    \label{fig:figure2}
    \medskip
    \small\raggedright \textit{
    The \textcolor{blue}{blue line} represents the average annual gross salary calculated using only the gross salary range [\$10,000 - \$80,000]. The \textcolor{orange}{orange dashed line} represents the average salary of the total number of employees in the database. The values of \textcolor{gray}{the bars} represent the value of the GDP (€ trillion). The \textcolor{PineGreen}{green dotted line} represents the number of employees by year (tens of millions).}
\end{figure}

The average gross annual salary is calculated for the entire population (Mean Gross Salary) and a calculation of the average gross annual salary for the range [\$10,000 - \$80,000] (Mean Gross Salary Range) is added, i.e., for each year, salaries and employees with gross annual salaries below \$10,000 and above \$80,000 are not considered. Both lines are highly correlated with each other, as well as with GDP in nominal euros, \textbf{Figure \ref{fig:figure2}}. Two periods of economic growth that significantly increase the average wage are characterized: 2001-2008 and 2015-2021. 

\subsection{Unemployment analysis}

\begin{figure}[H]
    \centering
    \caption{Unemployment trends by age and minimum wage 2002-2023}
    \includegraphics[width = 18cm]{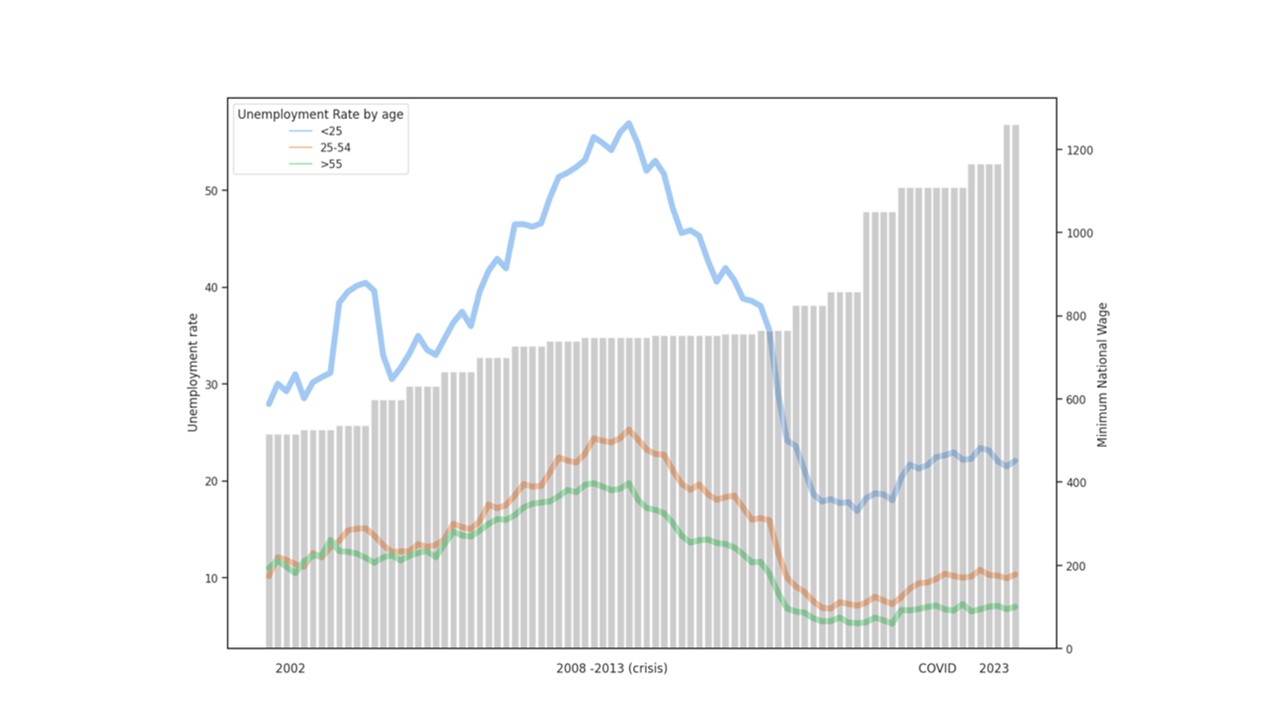}
    \label{fig:figure3}
    \medskip
    \small\raggedright \textit{
    The lines represent \textbf{the quarterly value of unemployment} according to the age of the worker, as collected by the National Statistics Institute (INE). The bars represent the minimum wage as defined by the government.}
\end{figure}

\textbf{Figure \ref{fig:figure3}} provides a visual analysis of the intuitive relationship between continuous increases in the minimum wage, especially in recent years, and unemployment. It is observed that youth unemployment has the highest elasticity for periods of economic crisis. No increase in unemployment among workers over 55 years of age is observed to be associated with an increase in the minimum wage. The National Statistics Institute (INE) periodically publishes unemployment figures. Quarterly figures are used in this case.

\subsection{Graph Analysis}

The following variables are evaluated.

\begin{itemize}
  \item Diff.- The difference between the average gross salary of all employees and the gross salary range [$10,000 - $80,000].
  \item Employees. - Number of employees by year.
  \item GINI. - Net Wage Gini Index.
  \item TAXES. - Withholding tax.
  \item GDP. - Gross Domestic Product.
  \item UnempRate. - Unemployment Rate.
  \item NMW. - National Minimum Wage.
  \item Mean Salary. - Mean Gross wage in Range.
  \item DEBT. - National Debt.
\end{itemize}

\begin{figure}[H]
    \centering
    \caption{Network visualization. }
    \includegraphics[width = 18cm]{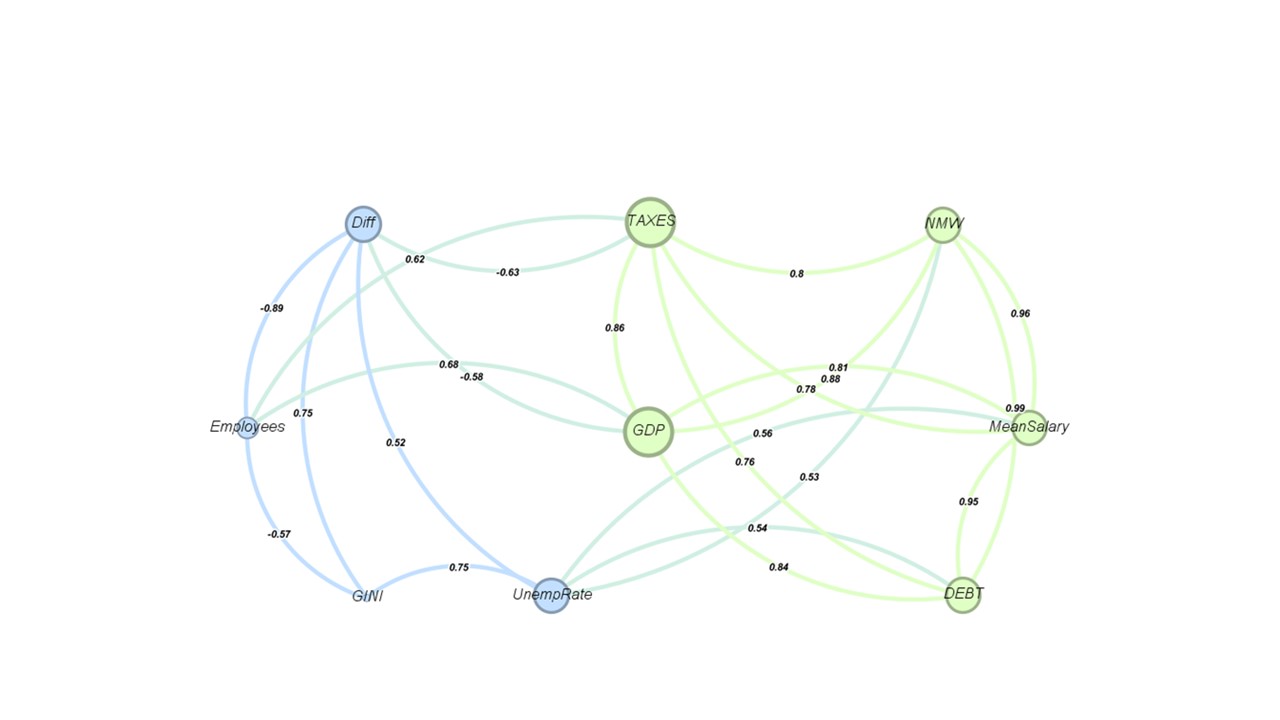}
    \label{fig:efigure4}
    \medskip
    \small\raggedright \textit{
    Colors represent communities detected in the graph. The \textbf{size of nodes} is proportional to their degree. \textbf{Each edge} represents a linear Spearman correlation, and the value of the correlation is displayed.}
\end{figure}

The size of the nodes depends on their degree. The higher the degree, the more the variable is related to the rest. The linear correlation between two nodes $\left(i,j\right)$ is represented by $corr\left(i,j\right)$. An $edge\left(i,j\right)$  is defined if $abs\left(corr\left(i,j\right)\right)\geq\ 0.5$. The linear Spearman correlation of each edge is given by its value. The color is defined by the two communities found. The interpretation of the graph shows that the variables of the same color have a stronger relationship with each other. Taxes and GDP have a crucial importance in a relationship with both community variables. GINI is only related to its community variables. The results are shown in \textbf{Figure \ref{fig:efigure4}}:

\begin{itemize}
    \item Modularity: 0,156
    \item Number of Communities: 2
    \item Average Degree: 4,889
    \item Maximum Nodes Degree: 6 (GPD, TAXES).
    \item Minimum Nodes Degree: 3 (GINI).
    \item Maximum linear correlation > 0.9: {(NMW - DEBT), (NMW - Mean Salary), (DEBT - Mean Salary)}.
\end{itemize}

Diff is the only node that has multiple negative correlations (Employees, GDP, and TAXES). It is strongly negatively correlated with the number of employees ($\rho = -0.89$). The NMW is strongly positively correlated with Mean Salary ($\rho = 0.96$). 

\subsection{Regression Analysis}

It is convenient to detail the results through an analytical analysis after a visual analysis of the study. Three types of regressions are analyzed: a multivariate linear regression, a regression model based on machine learning algorithms, and a time series regression model.
\subsubsection{Multiple linear regression}

The P-values and coefficients show which of the relationships in your model are statistically significant, and the nature of those relationships. Whether these relationships are statistically significant is indicated by the p-values for the coefficients. To develop a model that maximizes the R-squared value and eliminates the variables that lack statistical significance, different combinations were examined. The results are shown in  \textbf{Table \ref{table:table2}}. The GINI index of net salaries was used as the dependent variable. The objective is to analyze how the different macroeconomic variables studied will affect income inequality in Spain between 2001 and 2021.

\begin{table}[ht] 
    \caption{Ordinary Least Squares regression results}
    \centering
    \begin{tabular}{lllllll}\\
    \toprule
Variable      & coeff              & std error         & t      & P\textgreater{}|t| & {[}0.025   & 0.975{]}   \\
\hline \centering
const         & 0.111              & 0.021             & 5.37   & \textless{}0.001   & 0.066      & 0.156      \\
Employees     & 7.6 e-09           & 9.3 e-10          & 8.19   & \textless{}0.001   & 5.6 e-9    & 9.7 e-9    \\
DEBT          & 8.99 e-9           & 1.6 e-9           & 5.62   & \textless{}0.001   & 5.54 e-09  & 1.24 e-08  \\
Mean Salary R & 1.03 e-05          & 5.11 e-07         & 20.331 & \textless{}0.001   & 9.29 e-06  & 1.15 e-05  \\
MeanSalary    & -3.5 e-06          & 6.06 e-07         & -5.818 & \textless{}0.001   & -4.84 e-06 & -2.22 e-06 \\
GDP           & -5.08 e-08         & 9.5e-09           & -5.357 & \textless{}0.001   & -7.14 e-08 & -3.04 e-08 \\
Diff          & 1.39 e-05          & 6.11e-07          & 22.792 & \textless{}0.001   & 1.26 e-05  & 1.52 e-05  \\
ICP           & 0.0005             & 0.000             & 3.385  & 0.005              & 0.000      & 0.001      \\
TAXES         & -8.9 e-13          & 1.92 e-13         & -4.659 & \textless{}0.001   & -1.31 e-12 & -4.81 e-13 \\
\hline
              & \multicolumn{2}{l}{Dependent Variable} &        & Gini Index         &            &            \\
              & \multicolumn{2}{l}{R-Squared}          &        & 0.993              &            &            \\
              & \multicolumn{2}{l}{Durbin-Watson test} &        & 2.103              &            &            \\
    \bottomrule
    \end{tabular}
    \\

\small\raggedright
    \textit{\\
Variables meaning: \textbf{const} is the constant term in regression analysis is the value at which the regression line crosses the y-axis. \textbf{Employees} is the number of employees by year. \textbf{DEBT} is the national debt. \textbf{Mean Salary R} is the Mean gross salary range [\$10,000 - \$80,000]. \textbf{Mean Salary} is the Gross Salary. \textbf{Diff} is the difference between Mean Salary R - Mean Salary. \textbf{GDP} is the Gross Domestic Product. \textbf{ICP} is the Index of Consumer Price. \textbf{TAXES} is the total tax withheld during the year. \textbf{coeff} is the value of the coefficient that represents the average change in the response for a change of one unit in the predictor variable. \textbf{t} is the t-test statistic. Confidence Interval of Coefficients for $\alpha= 0.05$. }
\label{table:table2}
\end{table}

\subsubsection{Random Forest Regressor}

The preparation of the model run included the following steps:
\begin{enumerate}
    \item Define the dependent variable, here the Gini index.
    \item Parameterize the model by minimizing the MSE.
    \item Optimal model implementation.
\end{enumerate}
The mean square error was 0.0039. The model was optimized. Feature importance refers to a class of techniques for assigning scores to input features in a prediction model, indicating the relative importance of each feature in making a prediction. Relative scores can be used as a basis for subsequent analysis, highlighting which features are most relevant to the target. This information characterizes the model used to make predictions. Low scores suggest eliminating the variable to reduce the dimensionality of the problem. High scores suggest further analysis in more detail to be predictive for the model. The feature importance is shown in \textbf{Table \ref{table:table3}}.

\begin{figure}[H]
    \centering
    \caption{Permutation feature importance.}
    \includegraphics[width = 18cm]{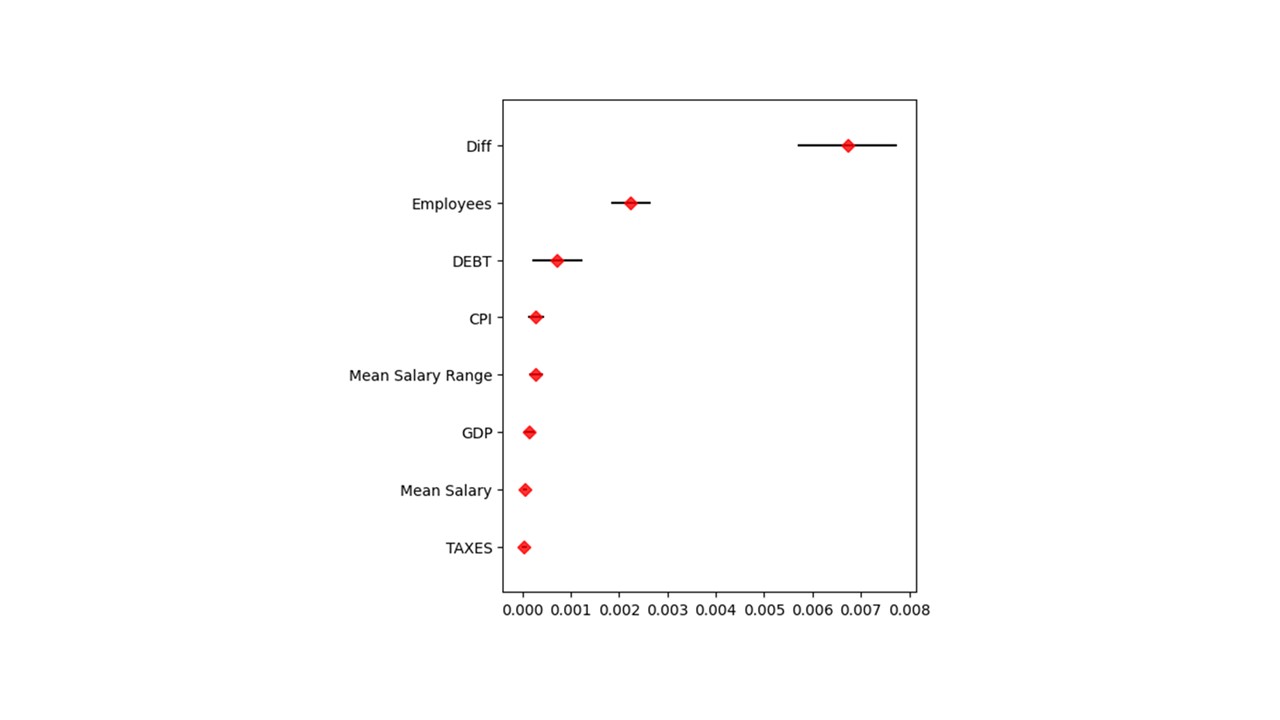}
    \label{fig:figure5}
    \medskip
    \small\raggedright \textit{
    The feature importance of the estimators for the dataset is calculated by the permutation importance function. The number of times a feature is randomly shuffled, and a sample of feature importance’s is returned. After 10 repetitions, the results stayed the same. The red dot represents the mean of the replicates. The line represents the standard deviation of each value. }
\end{figure}

\begin{table}[ht]
    \caption{Random Forest Regressor Model feature importance}
    \centering
    \begin{tabular}{lllllll}\\
    \toprule
Variable      & coeff              & std error         & t      & P\textgreater{}|t| & {[}0.025   & 0.975{]}   \\
const         & 0.111              & 0.021             & 5.37   & \textless{}0.001   & 0.066      & 0.156      \\
Employees     & 7.6 e-09           & 9.3 e-10          & 8.19   & \textless{}0.001   & 5.6 e-9    & 9.7 e-9    \\
DEBT          & 8.99 e-9           & 1.6 e-9           & 5.62   & \textless{}0.001   & 5.54 e-09  & 1.24 e-08  \\
Mean Salary R & 1.03 e-05          & 5.11 e-07         & 20.331 & \textless{}0.001   & 9.29 e-06  & 1.15 e-05  \\
MeanSalary    & -3.5 e-06          & 6.06 e-07         & -5.818 & \textless{}0.001   & -4.84 e-06 & -2.22 e-06 \\
GDP           & -5.08 e-08         & 9.5e-09           & -5.357 & \textless{}0.001   & -7.14 e-08 & -3.04 e-08 \\
Diff          & 1.39 e-05          & 6.11e-07          & 22.792 & \textless{}0.001   & 1.26 e-05  & 1.52 e-05  \\
ICP           & 0.0005             & 0.000             & 3.385  & 0.005              & 0.000      & 0.001      \\
TAXES         & -8.9 e-13          & 1.92 e-13         & -4.659 & \textless{}0.001   & -1.31 e-12 & -4.81 e-13 \\
              & \multicolumn{2}{l}{Dependent Variable} &        & Gini Index         &            &            \\
              & \multicolumn{2}{l}{R-Squared}          &        & 0.993              &            &            \\
              & \multicolumn{2}{l}{Durbin-Watson test} &        & 2.103              &            &            \\
    \bottomrule
    \end{tabular}

    \small\raggedright
    \textit{\\
                      Predictor. - Independence variables of Random Forest regressor model, and 
                     Gini index as dependence variable. Importance. - Value calculated.
 }
\label{table:table3}   
\end{table}
There is a second method of assessing importance in the regressor model called importance by permutation. This involves randomly shuffling each feature and calculating how the model performs. The features that have the greatest impact on performance are the most important ones. The importance of the permutation does not reflect the intrinsic predictive value of a feature per se, but rather the importance of that feature for a particular model. The result is shown in \textbf{Figure \ref{fig:figure5}}.

\subsection{Mean Salary Difference explained}

\begin{figure}[H]
    \centering
    \caption{Ratio of income to employees per band [2001-2016-2022].}
    \includegraphics[width = 18cm]{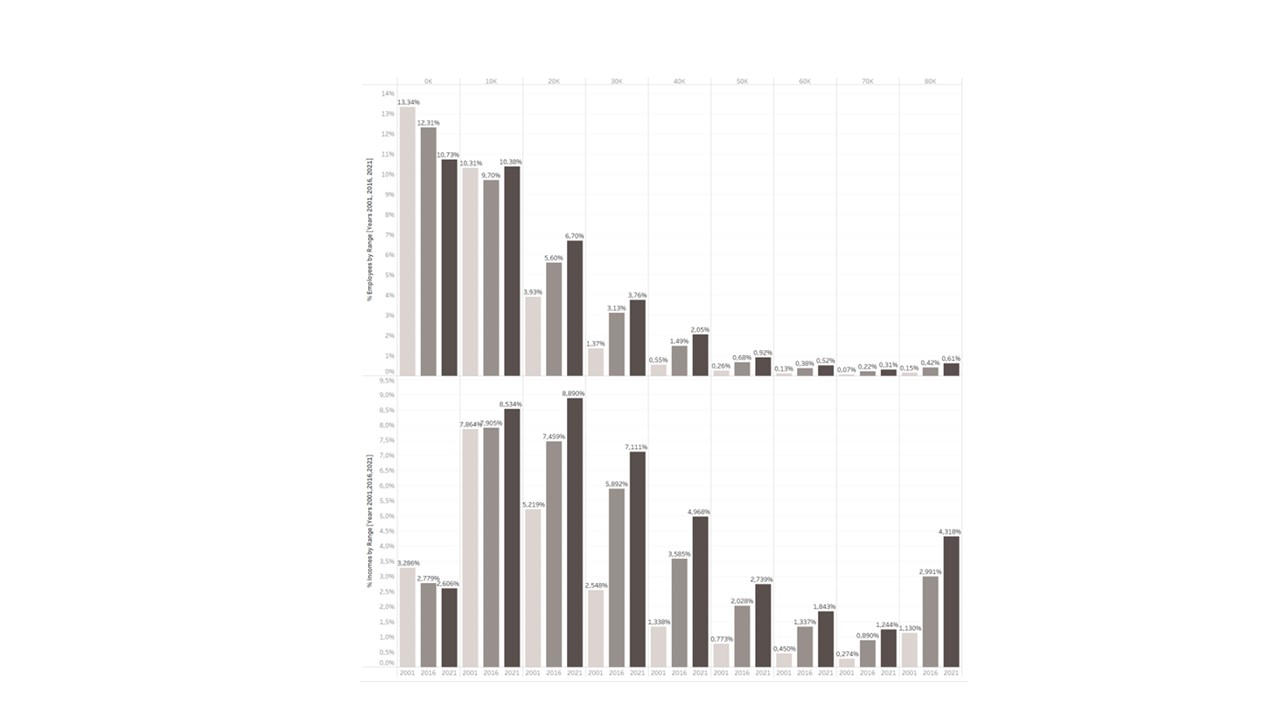}
    \label{fig:figure6}
    \medskip
    \small\raggedright \textit{
    The \textbf{share of employees} per band for the three years analyzed [2001-2016-2022] is shown in the upper part of the graph. The \textbf{share of the sum of their gross annual income per income} group is shown in the lower part of the graph.}
\end{figure}

The Mean Salary was calculated considering all employees in all bands of the analyzed database. The bands below €10,000 and above €80,000 have been excluded from the calculation of the Mean Salary Range. The first band corresponds to employees who did not work the calendar year and distorts the average. Similarly, the band above €80,000 is open and includes remarkably high salaries, which also distorts the average. For this reason, the variable 'average salary range' corresponds to the average of employees whose gross annual salary is between [10K-80K]. There has been a shift in the proportions of employees by rank, particularly since 2016, coinciding with more substantial increases in the minimum wage. Over the years, the [20K-40K] range has increased from 5.85\% (12.51\% of gross income) of the population in 2001 to 15.61\% (20.96\% of gross income) in 2021. This is because underemployment (workers who want to work more hours and whose contracts do not cover 40 hours a week or all months of the year) exceeds 15\% of total employment (Brecha de género en el empleo por tipo de empleo y periodo, n.d.). This analysis is shown in \textbf{Figure \ref{fig:figure6}}.

Spearman's linear correlation value between the DIFF variable and GINI was $\rho=\ 0.75$  shown in Fig 4. DIFF has been found to be the most influential in predicting the Gini Index score. 

\begin{figure}[H]
    \centering
    \caption{Ratio of income to employees per band [2001-2016-2022].}
    \includegraphics[width = 18cm]{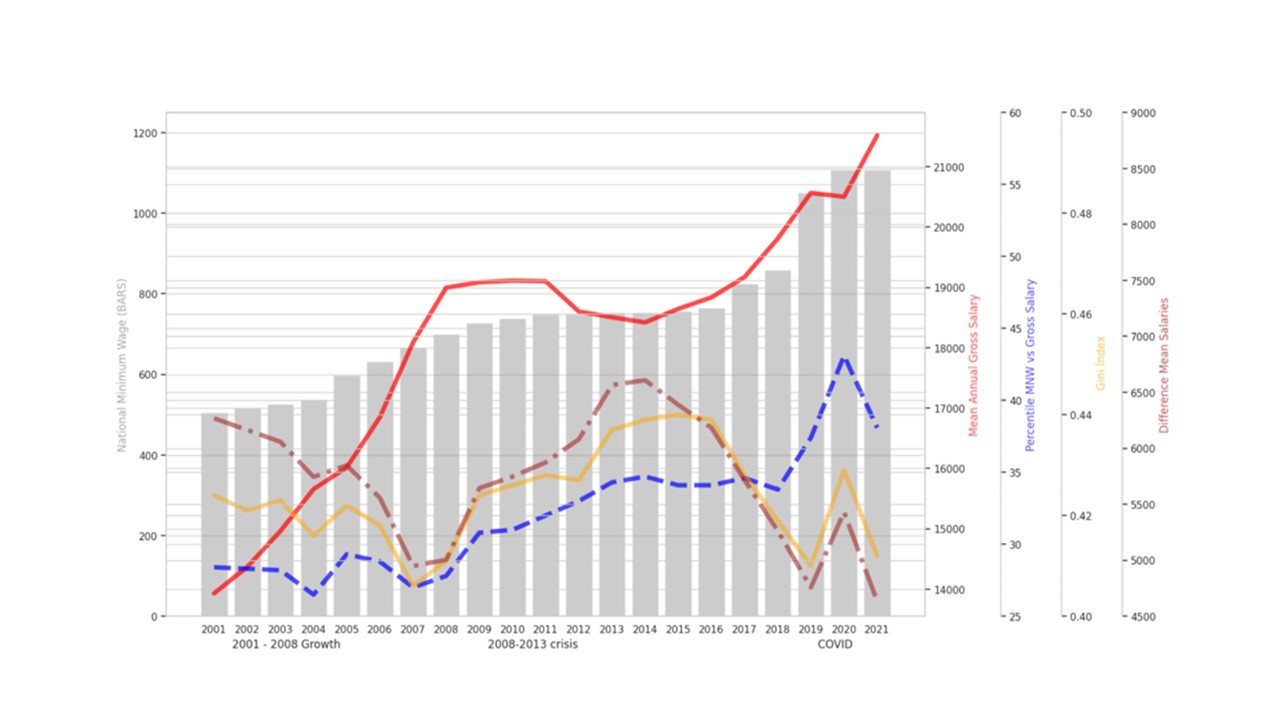}
    \label{fig:figure7}
    \medskip
    \small\raggedright \textit{
    For each year, the \textbf{difference} between the average gross salary of all employees and the gross salary range [$10,000 - $80,000] is evaluated and represented by \textcolor{brown}{he brown line}. The \textbf{National Minimum Wage} is represented by \textcolor{gray}{bars} in the figure. The \textcolor{orange}{cream line} represents the Gini index corresponding to the net income values of the employees each year. The \textcolor{blue}{blue line} represents the evolution of the NMW \textbf{percentile} versus the annual gross salary vector.}
\end{figure}

The evolution of the percentile of the minimum wage over the period under study is analyzed and presented in \textbf{Figure \ref{fig:figure7}}, together with other macroeconomic variables. A percentile above 40 is noticeable. This means that 40\% of the wages recorded in the database are paid below the Minimum Wage. The evolution of the percentile is related to an increasing minimum wage and to the evolution of the average gross wage.

\section{Discussion}

This paper is based on the statistical analysis of macroeconomic time series published by government agencies of the Spanish government. The Bank of Spain, among others, predicted in 2017 that increasing the minimum wage would cause unemployment to rise \cite{Lacuesta_Gabarain2019-cg}. Experience has shown that none of these forecasts have been accurate. Fig 3 shows that an increase in the minimum wage is compatible with the creation of new jobs. The average wage increases, and income inequality becomes fairer. These results are consistent with some studies mentioned in the paper. In the United Kingdom (1999-2010), the development of wage inequality was examined. The impact on inequality of the introduction of the minimum wage in 1999 was moderate. This can be explained by the fact that the minimum wage was introduced at a level that was below the 10th percentile of the earnings distribution \cite{Stewart2002-fg}. However, the specific features of the structure of the Spanish economy must be considered. One reason for its impact on reducing income inequality may be that the most recent minimum wage is at the 30-40th percentile.

Extrapolating these results to other economies, or even to Spain in the near future, may be unwise. An increase in economic inequality can lead to a lack of trust in governing politicians \cite{Andersen2012-sx}.  In 2022, Piketty et al. suggest that policy discussions on inequality should focus on policies that affect pre-tax inequality, rather than focusing exclusively on tax redistribution (Predistribution vs. Redistribution: Evidence from France and the US, n.d.). This paper shows that redistribution does not change the dynamics of economic inequality. As the minimum wage increases, there is an increase in the number of workers in the \$20,000-\$40,000 range. This tends to reduce income inequality and raise the average salary. And this is true for the Spanish economy since the minimum wage is far from the level of the main European economies, such as France or Germany. There is a question as to the extent to which an increase in the minimum wage will have a negative impact on the economy. It could be argued that as it approaches that of Germany, either productivity increases or economic imbalances could arise, which have been widely analyzed. Inequality tends to be pro-cyclical. Low-income households and young people tend to be hit harder by recessions. The distribution of labor and capital income differs across countries. Therefore, the cyclicality of income inequality may also differ \cite{Clemens2020-tl}. 

Three different regression models were used. All of them confirm that the DIFF variable is the best predictor of income inequality. It is particularly noticeable when there is a movement towards annual salaries above €20,000, which is largely due to the increase in the minimum wage. There is reason to believe that the wage structure will tend to follow the same trend as the minimum wage increases, catching up with major economies, if the minimum wage percentile is above 20. In addition, income inequality will improve. This is measured by a decline in the Gini index. The collection of corporate income tax has risen to a record figure of more than 200 billion euros for the first time in 2019 (Recaudación y Estadísticas del Sistema Tributario Español: Ministerio de Hacienda y Función Pública, n.d.). Therefore, there is no evidence of a reduction in corporate profits because of a minimum wage increase.

\section{Conclusion}

The most important conclusion is that an increase in the minimum wage in the period analyzed has led to an increase in the wealth of the country, increasing employment and company profits, and is postulated, under the conditions analyzed, as an effective method for the redistribution of wealth.

Using the data provided by the AEAT, it could be analyzed databases that represent the entire population studied. More accurate results and more efficient predictive models are provided by these counts. The results offered in this paper should be analyzed in the context of the particularity of the Spanish economy.

Various machine-learning models show that income inequality has been reduced by raising the minimum wage. Raising the minimum wage has not led to inflation or increased unemployment. On the contrary, it has been consistent with increased net employment, contained prices, and increased corporate profit margins.

Other aspects, such as the profitability of movable and immovable assets and the income of the self-employed, should be included in a subsequent analysis.



\end{document}